\documentclass[prl, superscriptaddress,twocolumn]{revtex4}
\usepackage{epsfig,amsmath,amssymb,latexsym}
\usepackage{graphicx}
\usepackage{dcolumn}
\usepackage{bm}
\usepackage{subfigure}

\begin{document}

\title{Fine and Large Coulomb Diamonds in a Silicon Quantum Dot}

\author{T. Kodera}
\affiliation{Institute for Nano Quantum Information Electronics, the University of Tokyo, 
4-6-1 Komaba, Meguro-ku, Tokyo 153-8505, Japan}
\email{kodera.t.ac@m.titech.ac.jp}

\author{T. Ferrus}
\affiliation{Hitachi Cambridge Laboratory, J.J. Thompson Avenue, Cambridge CB3 0HE,  United Kingdom}

\author{T. Nakaoka}
\affiliation{Institute for Nano Quantum Information Electronics, the University of Tokyo, 
4-6-1 Komaba, Meguro-ku, Tokyo 153-8505, Japan}
\affiliation{Institute of Industrial Science, the University of Tokyo, 4-6-1 Komaba, Meguro-ku, Tokyo 153-8505, Japan}

\author{G. Podd}
\affiliation{Hitachi Cambridge Laboratory, J.J. Thompson Avenue, Cambridge CB3 0HE,  United Kingdom}

\author{M. Tanner}
\affiliation{Hitachi Cambridge Laboratory, J.J. Thompson Avenue, Cambridge CB3 0HE,  United Kingdom}

\author{D. Williams}
\affiliation{Hitachi Cambridge Laboratory, J.J. Thompson Avenue, Cambridge CB3 0HE,  United Kingdom}

\author{Y. Arakawa}
\affiliation{Institute for Nano Quantum Information Electronics, the University of Tokyo, 
4-6-1 Komaba, Meguro-ku, Tokyo 153-8505, Japan}
\affiliation{Institute of Industrial Science, the University of Tokyo, 4-6-1 Komaba, Meguro-ku, Tokyo 153-8505, Japan}
\affiliation{Research Center for Advanced Science and Technology, the University of Tokyo, 
4-6-1 Komaba, Meguro-ku, Tokyo 153-8505, Japan}

\date{\today}

\begin{abstract}
We experimentally study the transport properties of silicon quantum dots (QDs) fabricated from a highly doped n-type silicon-on-insulator wafer. Low noise electrical measurements using a low temperature complementary metal-oxide-semiconductor (LTCMOS) amplifier are performed at 4.2 K in liquid helium. Two series of Coulomb peaks are observed: long-period oscillations and fine structures, and both of them show clear source drain voltage dependence. We also observe two series of Coulomb diamonds having different periodicity. The obtained experimental results are well reproduced by a master equation analysis using a model of double QDs coupled in parallel.
\end{abstract}

\maketitle

Single-electrons in quantum dots (QDs) have been proposed as one of the suitable candidates for realizing solid-state quantum bits (qubits).~\cite{Loss,Nielsen} Recently, the essential requirements of controlling and measuring single electron states have been achieved in GaAs-based semiconductor QDs.~\cite{Johnson,Petta,Koppens,Wiel} On the other hand, Si-based QDs have attracted interest because a long electron-spin coherence time is expected owing to the small spin-orbit coupling in addition to the almost spin-zero nuclear background.~\cite{Tahan} Despite the interest, the single-electron states have been less investigated in Si QDs~\cite{Gorman,Takahashi,Fujiwara1,Fujiwara2} because the relatively heavy effective mass makes the manipulation of single electrons technically challenging. The larger effective mass of electron in Si than GaAs makes the confinement potential energy smaller. To obtain quantum confined electronic states, it is necessary to fabricate much smaller nanodevices. In addition, it is more difficult to make a clean two-dimensional electron gas in Si than GaAs and to avoid noise from the device itself because of the interface effect and impurities. 

In order to probe the single-electron states in such a Si system, it is important to measure nanodevices in a low-noise system. Measurement devices such as a transimpedance or charge integrator amplifier are more suitable for use in a low temperature system at 4.2 K owing to the variation of noise gain with temperature. 

In a nanoscale device, the confining potential structure might become complex. Therefore, it is important to perform a theoretical simulation in which a model that can reproduce the obtained experimental results is used. From the simulation we can extract the potential structure of the device and obtain data that can be fed back into the fabrication process. 

In this work, we experimentally study the transport properties of Si QDs fabricated from a highly doped n-type Si-on-insulator (SOI) wafer. Low-noise electrical measurements using a low-temperature complementary metal-oxide-semiconductor (LTCMOS) amplifier are performed at 4.2 K in liquid helium.~\cite{Hasko} We successfully observed clear Coulomb oscillations and diamond-shaped Coulomb blockade regions. In particular, we observed a peculiar pattern of a large diamond containing several small diamonds in it. We calculate the transport properties by solving master equations for several QD systems to reproduce the obtained features. The results provide the configuration, size, and charging energy of QDs in the device and allow us to propose a mechanism for the formation of a double QD inside a single island patterned by electron-beam (EB) lithography.

\begin{figure}[tb]
\begin{center}
\includegraphics[width=8cm]{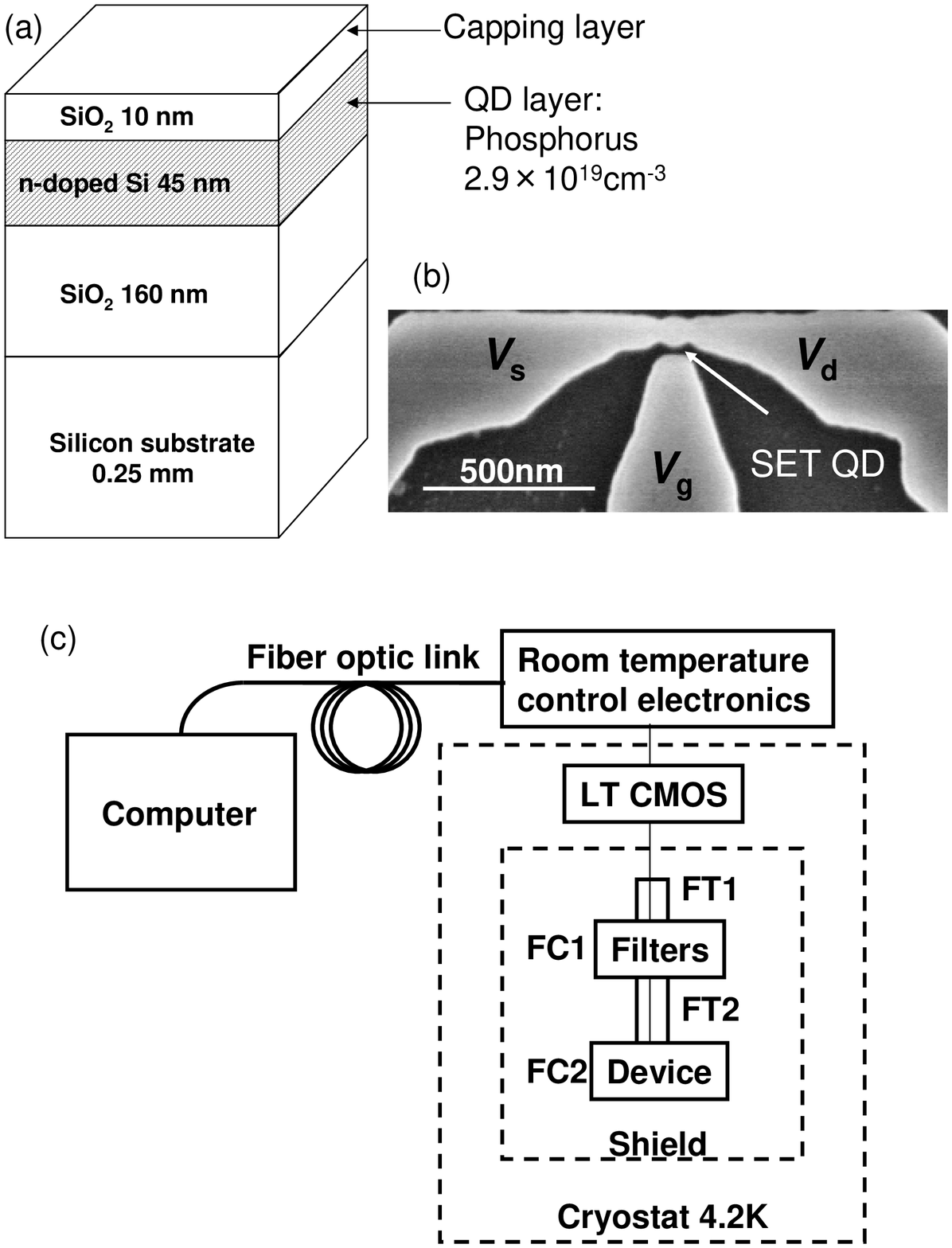}
\end{center}
\caption{ (a) Schematic cross section of the device. The material and thickness of each layer are indicated. (b) SEM image of the device. The diameter of the SET QD patterned by EB lithography (indicated by an arrow) is $\sim$80 nm. (c) Schematic diagram of the measurement setup.}
\label{f1}
\end{figure}

Figure \ref{f1}(a) shows the wafer profile with phosphorus doping in Si to 2.9$\times$10$^{19}$ cm$^{-3}$. 
By introducing a 10 nm capping SiO$_2$ layer we can avoid damage to the Si device layer due to contaminants and impurities during the fabrication process. Introducing a 45 nm n-doped Si layer between SiO$_2$ layers also provides strong vertical electrical confinement. 
The QD structures are fabricated using EB lithography and reactive ion etching to achieve trench isolation. Figure \ref{f1}(b) shows a scanning electron microscope (SEM) image of the device. Bright (dark) regions correspond to the electrodes and QDs (the SOI). The diameter of the patterned QD region is about 80 nm. The electron wavefunction is confined to below 50 nm using a postoxidation technique. The current $I_{sd}$ through the QD in the single-electron transistor (SET) structure is measured when a voltage $V_{sd}$ is applied between the source ($V_{s}$) and drain ($V_{d}$) contacts. By applying a gate voltage $V_{g}$, the electrochemical potential of the QD can be modulated. 

The measurement of the current through the SET must be accurate to pA or better to observe fine features. The measurement probe consists of a custom LTCMOS integrated circuit that provides various voltages to the devices and that contains the measurement circuit as shown in Fig. \ref{f1}(c).~\cite{Hasko} Communication to and from the computer is realized through an optical high-speed fiber link connected to a room-temperature data acquisition and communication box. The SET current is measured using a charge integrator amplifier circuit where the signal current is integrated onto 1 or 10 pF capacitors, making the integration period as short as 10-20 $\mu$s. All lines are filtered by a single-stage resistor-inductor-capacitor low-pass filter having a cutoff frequency of about 80 kHz at 4.2 K. Both the filters and the devices are further shielded from parasitic electrical noise by Faraday cages (FCs) using high-resistance ferrite bead feedthroughs (FTs) for input and output cabling. This arrangement efficiently suppresses the radiation and conduction of high-frequency noise.

\begin{figure}[tb]
\begin{center}
\includegraphics[width=8cm]{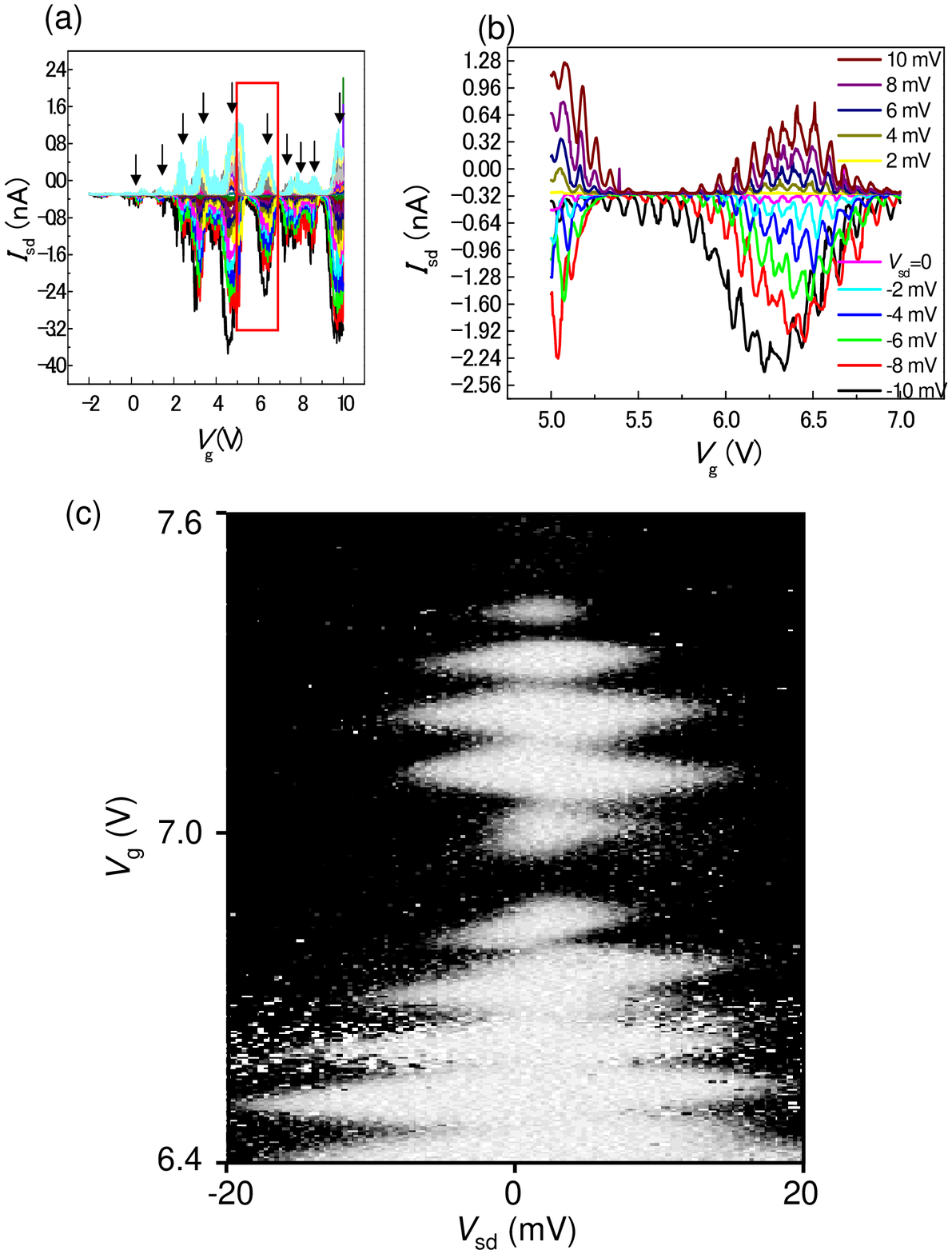}
\end{center}
\caption{ (Color online) (a) Measured $I_{sd}$ vs. $V_{g}$ for various $V_{sd}$. $V_{sd}$ is changed from -10 to 10 mV in 0.5 mV steps. Coulomb oscillation peaks are indicated by arrows. (b) Enlargement of measurement plot for the region of $V_{g}$ defined by the red square in (a). The values of $V_{sd}$ for each plot are indicated. (c) Gray-scale plot of the Coulomb diamonds or differential conductance d$I_{sd}$/d$V_{sd}$ in the plane of $V_{sd}$ and $V_{g}$. The Coulomb blockade and conducting region are shown in white and black, respectively. There is an offset of +2.1 mV in $V_{sd}$ due to asymmetric barriers and the experimental setup. }
\label{f2}
\end{figure}

Figure \ref{f2}(a) shows the measured $V_{g}$ dependence of $I_{sd}$ at different $V_{sd}$. Ten Coulomb oscillations (indicated by arrows) are observed from $V_{g}$ = -2 to 10 V. The Coulomb peaks become smaller with decreasing $V_{sd}$ from 10 to 2 mV, change signs, and become increasingly negative with decreasing $V_{sd}$ from 2 to -10 mV. No oscillations are observed below $V_{g}=-2$ V, probably either because of the reduced conductivity between the QD and source (or drain) contact or because of the pinch-off in the QDs. Figure \ref{f2}(b) shows an enlargement of the measurement plot for the region of gate voltage defined by the red square in Fig. \ref{f2}(a). Fine structures are clearly observed on top of the Coulomb peaks that show the same $V_{sd}$ dependence as the peaks with long periods. These features have been confirmed by several measurements with different sweep rates and with very high resolution. They are still observable after thermally cycling the device between 4.2 K and room temperature and are present on similarly patterned devices. The two observed series of Coulomb oscillations are attributed to single-electron tunneling into the electronic state of the double QD incidentally created inside the SET QD region, which we shall explain in detail later. 

Figure \ref{f2}(c) shows the measured gray-scale plot of the Coulomb diamonds or differential conductance d$I_{sd}$/d$V_{sd}$ versus $V_{sd}$ and $V_{g}$. The Coulomb blockade and conducting region are shown in white and black, respectively. We observe clear diamond-shaped Coulomb blockade regions whose sizes are modulated as we change $V_{g}$. At $V_{g}$ = 6.9 and 7.5 V the Coulomb blockade diamond disappears. The diamonds near the conducting region have shorter lateral length along the $V_{sd}$ axis. Although the lateral length along $V_{sd}$ depends on $V_g$, the vertical length along $V_g$ or the periodicity of the Coulomb oscillation is almost constant. Two Coulomb diamond series with similar features have been recently observed and explained by a stochastic Coulomb blockade with asymmetric triple QDs coupled in series, fabricated by a pattern-dependent oxidation process.~\cite{Monoharan} However, it seems unlikely that QDs formed in the tunnel barrier in our device because of the patterned device dimensions and oxidation conditions.

Here we compare the experimental results with a theoretical calculation in order to study the topology of our QD system. We calculate the differential conductance in the framework of the modified orthodox model that includes the shell structure.~\cite{Tarucha} The tunneling current can be obtained for an arbitrarily connected double QD system capacitively coupled with a single gate.~\cite{Danilov} In our analysis we follow the solution of an orthodox model for a double QD and modify it to include the shell structure. 
The master equation can be solved numerically as follows: for any given ($V_{sd}$,$V_{g}$) we trace the evolution of the probability distribution until an equilibrium distribution is reached. The current can be computed for an arbitrary double QD system. We show the mathematical results for two types of double QD system coupled in series and in parallel.

\begin{figure}[tb]
\begin{center}
\includegraphics[width=8cm]{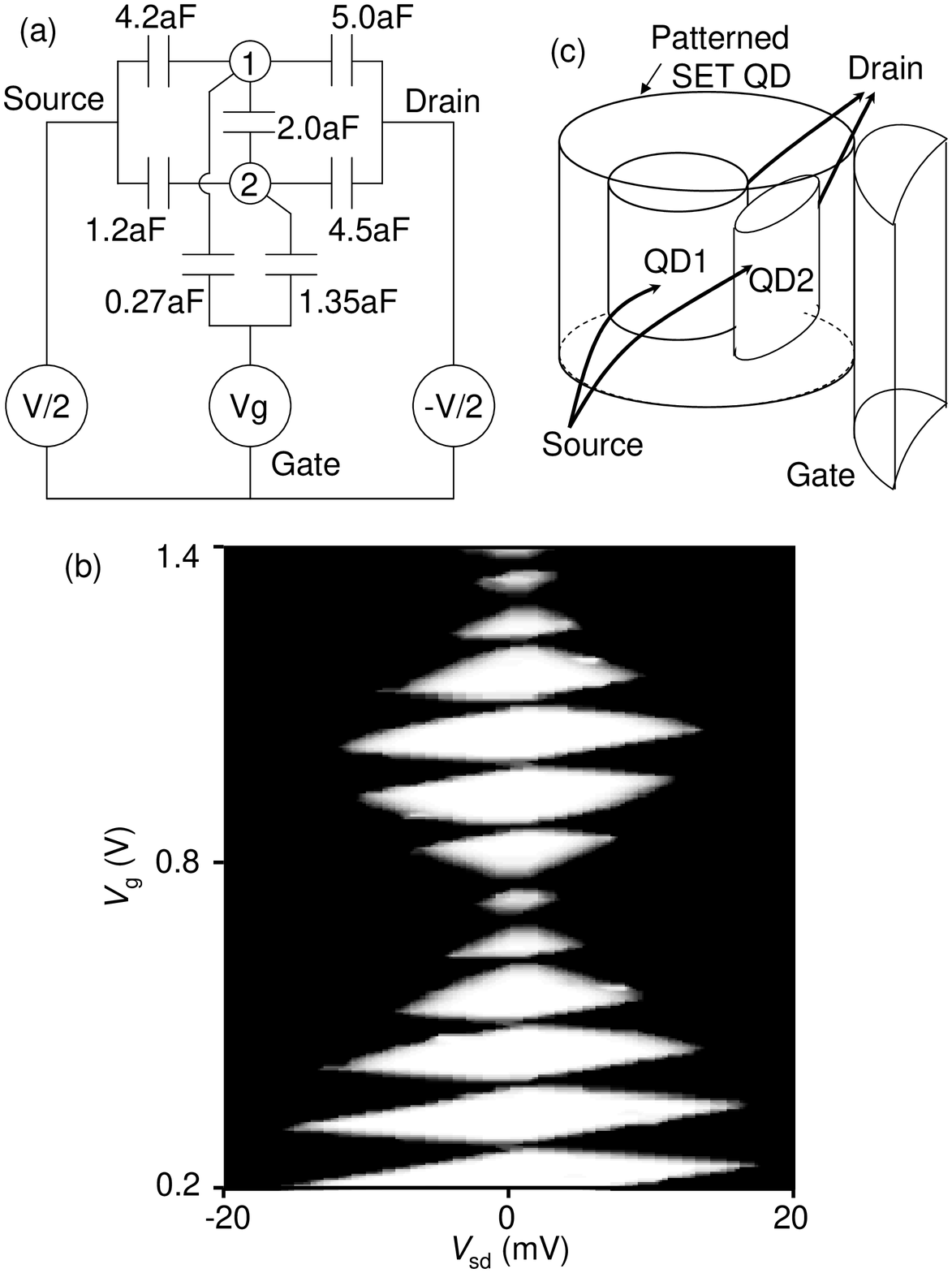}
\end{center}
\caption{ (a) Schematic circuit model of double QD coupled in parallel. The system parameters are indicated. (b)  Simulated differential conductance plot as a function of $V_{sd}$ and $V_{g}$. The Coulomb blockade and conducting region are shown in white and black, respectively. (c) Schematic diagram showing the possible configuration of QD1 and QD2 within the patterned SET QD.}
\label{f3}
\end{figure}


We have calculated the differential conductance for the model with double QDs coupled in parallel as shown in Fig. \ref{f3}(a). The coupling of one QD (QD2) with the gate electrode is five times stronger than that of the other QD (QD1). QD2 is smaller than QD1, thus the charging energy is larger. Here we introduce asymmetric capacitances between the source and the QDs and between the drain and the QDs in order to reproduce the experimental results. Inter-QD electrostatic coupling $C_{int}$ is 2 aF. 

The calculated results agrees well with the experimental ones as shown in Fig. \ref{f3}(b). In particular, the gradual change in the lateral size of the Coulomb diamonds is perfectly reproduced. This peculiar pair of series of Coulomb diamond patterns is due to parallel conduction through the QDs differently coupled to the gate electrode. Small diamond patterns appear in the larger diamond. Within the larger diamond (QD1), the number of electrons is fixed while electrons are added one by one to QD2. Outside the larger Coulomb blockade diamonds, the current always flows through QD1 even though QD2 is in the Coulomb blockade condition.
The diameters $d_{eff}$ of QD1 and QD2 are estimated as $\sim$28 and $\sim$22 nm, respectively, using $d_{eff}$$\sim$$C$/4$\epsilon_r$$\epsilon_0$ ($\epsilon_r$=11.7 in Si). Here $C$ is the sum of capacitances related to each QD. The charging energies $e^2$/$C$ of QD1 and QD2 is derived as $\sim$14 and $\sim$18 meV, respectively, where $e$ is a single electron charge. The relative position of the two QDs is estimated by calculating the distance $d$ between the two disklike QDs. By using $d$=$\epsilon_r$$\epsilon_0$$d_{eff}$$^2$/4$C_{int}$ for the inter-QD electrostatic coupling, we find that $d$$\sim$25 nm. According to this value, QD2 exists inside the patterned SET QD. Applied gate voltage is positive in the measurement; thus, a the some localized state is induced near the gate as shown in Fig. \ref{f3}(c). The thickness of the phosphorus-doped Si layer is 45 nm and the QD1 diameter is $\sim$28 nm, which is comparable to the thickness; thus, the three-dimensional confinement and coupling may have to be considered. 

\begin{figure}[tb]
\begin{center}
\includegraphics[width=8cm]{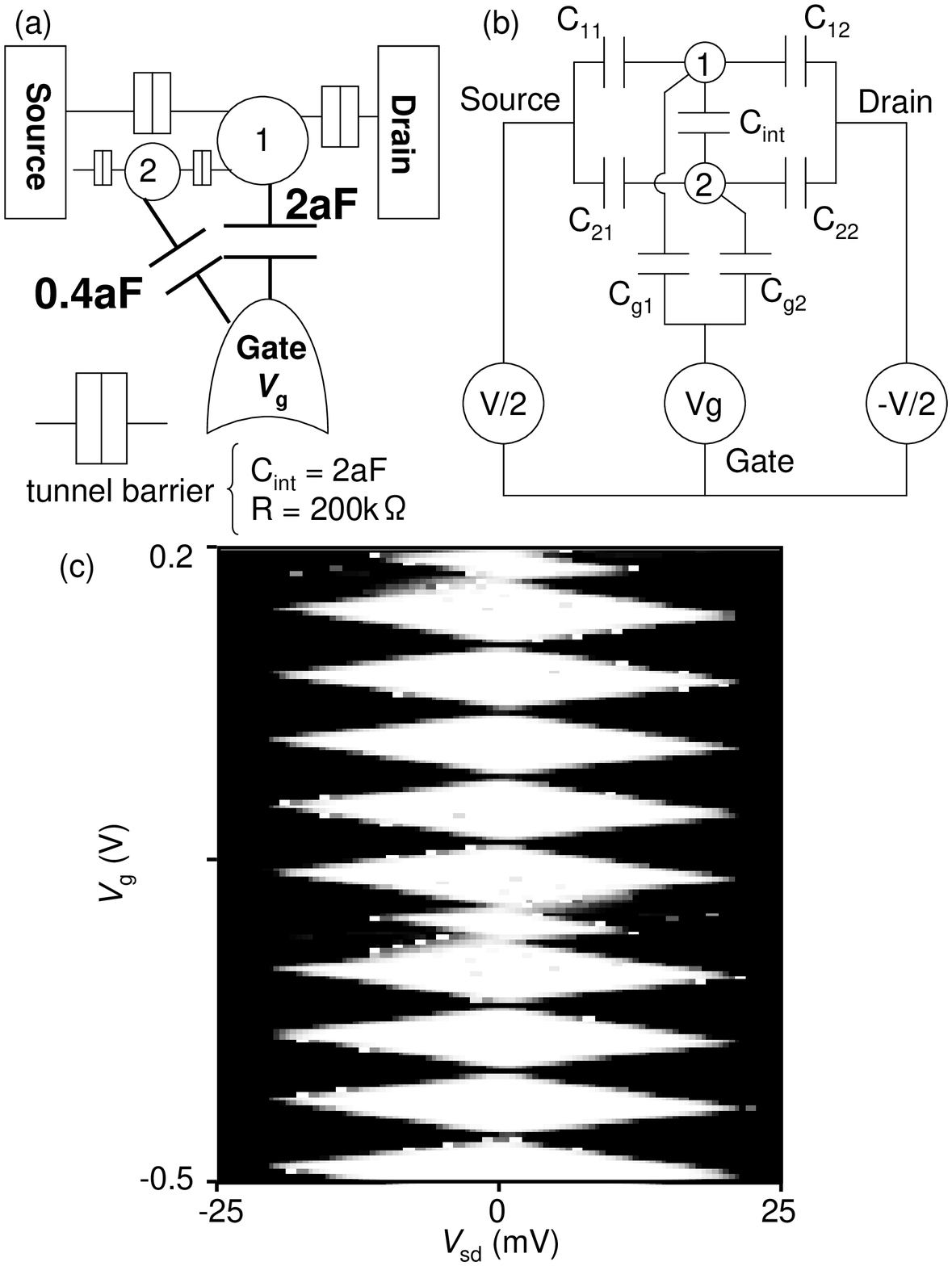}
\end{center}
\caption{ (a) Schematic model of a double QD coupled in series. (b) Schematic circuit model of an arbitrary double QD. System parameters $C_{11}$=$C_{12}$=$C_{21}$=$C_{int}$=2$\times$10$^{-18}$ F, $C_{22}$=0, $C_{g1}$=2 $\times$10$^{-18}$ F, and $C_{g2}$=0.4$\times$10$^{-18}$ F are applied. (c) Simulated differential conductance plot as a function of $V_{sd}$ and $V_{g}$. }
\label{f4}
\end{figure}


Here we note that other QD geometries cannot reproduce our experimental data. For example, as a representative case, we show the calculated result
for two QDs coupled in series as shown in Fig. \ref{f4}(a). This type of QD system was used in ref. \cite{Monoharan} to  explain the two series of Coulomb diamond patterns as shown in Fig. \ref{f2}(c). Their device, which was fabricated using a pattern-dependent oxidation process,
was modeled using three different-size QDs coupled in series. 
Here we consider a double QD system in series with a small incidental QD (QD2) at the barrier between the source contact and a larger main QD (QD1). 
The electrostatic coupling of QD2 with the gate electrode is smaller than that of QD1. In the arbitrary double QD model shown in Fig. \ref{f4}(b), we substitute the system parameters of $C_{11}$=$C_{12}$=$C_{21}$=$C_{int}$=2$\times$10$^{-18}$ F, $C_{22}$=0, $C_{g1}$=2 $\times$10$^{-18}$ F, and $C_{g2}$=0.4$\times$10$^{-18}$ F. 

The calculated differential conductance is shown in Fig. \ref{f4}(c). A modulation is observed in the lateral size of the Coulomb diamonds. This is due to the modification of the potential of QD1 by the inter-QD Coulomb energy when the number of electrons in QD2 is changed.
Using the series model, there is no mechanism to produce the gradual change in size of the Coulomb diamonds. Therefore, we conclude that it is difficult to reproduce our experimental results using a series model. 

We experimentally study the transport properties of a Si QD fabricated from an SOI wafer at 4.2 K. Two series of Coulomb diamonds having different periodicity are observed. In order to study the device topology, a master equation is solved for an arbitrary double QD system in which where the shell structure is taken into account. We find that the experimental data are well reproduced by a parallel rather than a series QD model. 

Although the formation mechanism of a second QD near the gate electrode is still not clear at present, recent experimental evidence suggests that it may be related to the presence of localized states and disorder in the device. To this end, further analysis is necessary and three-dimensional coupling with the patterned QD should be considered.

In order to realize disklike QDs in a few-electron regime, it is desirable to reduce the thickness of the Si device layer, to fabricate smaller QDs, and to attach a top gate electrode that would provide flexibility in tuning the QD potential. These may lead to understanding of the origin of the parallel QD. This is of importance because such a SET is being considered as a readout architecture for quantum computation.

The authors thank Mr. G. Yamahata for fruitful discussions. This work was financially supported by EPSRC-UK (GR/S24275/01 and GR/S15808/01) and the Special Coordination Funds for Promoting Science and Technology in Japan.

\end{document}